\documentclass[prl]{revtex4}
\usepackage{bm}
\usepackage{epsfig}
\usepackage{amsmath}
\usepackage[pdftex,ps2pdf,bookmarks=true,bookmarksnumbered=true,breaklinks=true,hypertexnames=false,linkbordercolor={0 0 1},pdfborder={0 0 112.0}]{hyperref}
\begin{document}

\title{Curve crossing induced dissociation : An analytically solvable model}
\author{Aniruddha Chakraborty}
\affiliation{School of Basic Sciences, Indian Institute of Technology Mandi,\\
Mandi, Himachal Pradesh, 750001, India.}
\date{\today}

\begin{abstract}
\noindent In our earlier papers we have proposed an analytically solvable model for the two state curve crossing problem which assumes the coupling to be a Dirac delta function. It is used to calculate the effect of curve crossing on electronic absorption spectrum and Resonance Raman excitation profile for the case of harmonic potentials. In this paper we have extended our model to deal with the curve crossing induced dissociation cases. Our method is used in this paper to calculate the effect of curve crossing induced dissociation on electronic absorption spectrum and Resonance Raman excitation profile. In this paper, a model consisting of a Harmonic oscillator  and a Morse oscillator, coupled by Dirac delta function, is solved.
\end{abstract}

\maketitle

\section{Introduction}

\noindent Nonadiabatic transition due to potential curve crossing is one of the most important mechanisms to effectively induce electronic transitions in collisions \cite{Naka1,Naka2,R1,R2,R3,R4,R5,AniThesis,AniBook, AniRev,Ani1,Ani2,Ani3}. This is a very interdisciplinary concept and appears in various fields of physics, chemistry and biology \cite{Naka2}. Two state curve crossing is in general classified into the following two cases according to the crossing scheme: (1) Landau-Zener (L.Z) case, in which the two diabatic potential curves have the same signs for the slopes and (2) non-adiabatic tunnelling (N.T) case, in which the diabatic curves have the opposite sign for slopes. There is also a
non-crossing non-adiabatic transition called the Rosen-Zener-Demkov type \cite{Naka1,Naka2}, in which two adiabatic potentials are in near resonance at large $R$. The theory of non-adiabatic transitions dates back to $1932$, when the
pioneering works for curve-crossing and non-crossing were published by Landau \cite{Landau}, Zener \cite{Zener} and Stueckelberg \cite{Stueckelberg} and by Rosen and Zener \cite{Rosen} respectively. Since then numerous papers by many
authors have been devoted to these subjects, especially to curve crossing problems\cite{Naka1,Naka2}. In our earlier papers we have proposed an exactly solvable model for the two state curve crossing problem which assumes the coupling to be a Dirac delta function \cite{AniThesis,AniBook,AniRev,Ani1}. This model is used to calculate the effect of curve crossing on electronic absorption spectrum and on Resonance Raman excitation profile for the case of harmonic potentials \cite{AniThesis,AniBook,AniRev,Ani1}. We have later generalized our model to deal with general multi-channel curve crossing problem too \cite{Ani2}. Even very recently our model ia extended to deal with nonadiabatic tunneling in an ideal one dimensional semi-infinite periodic potential systems \cite{Ani3}. We have also proposed an analytical method for the two state curve crossing problem for any coupling \cite{Ani4}. We have also used our analytically  solvable to deal with scattering problems \cite{Ani5}. The same method has been applied recently to the case of predissociation \cite{Ani6}. Our work is in progress to deal with nonadiabatic tunneling in an ideal one dimensional finite periodic potential systems \cite{Ani7}.  In this paper we have extended our model to deal with the curve crossing induced dissociation cases. In this paper we analyze the effect of curve crossing induced dissociation on electronic absorption spectrum and Resonance Raman excitation profile using a model consisting of a Harmonic oscillator and a Morse oscillator, coupled by Dirac delta function.

\section{The model}

\noindent We consider two diabatic curves, crossing each other. There is a coupling between the two curves, which causes transitions from one curve to another. This transition would occur in the vicinity of the crossing point. In particular, it will occur in a narrow range of $x$, given by
\begin{equation}
\label{1}\left|V_1(x)-V_2(x)\right|\simeq \left|V(x_c)\right|.
\end{equation}
where $x$ denotes the nuclear coordinate and $x_c$ is the crossing point. $V_1$ and $V_2$ are the diabatic potentials and $V$ represent the coupling between them. In reality the transition between $V_{1}(x)$ and $V_{2}(x)$ occur most effectively at the crossing, because the necessary energy transfer between the electronic and nuclear degrees of freedom is minimum there. Therefore it is interesting to analyze a model, where coupling is localized in space near $x_c$ rather than using a model where coupling is same everywhere (i.e. constant coupling). Thus we put
\begin{equation}
\label{2}V(x)=K_0\delta (x-x_c),
\end{equation}
here $K_0$ is a constant. This model has the advantage that it can be analytically solved \cite{AniThesis, AniBook, AniRev,Ani1,Ani2,Ani3,Ani4,Ani5,Ani6}.

\section{Exact analytical solution}

\noindent We start with a particle moving on any of the two diabatic curves. The problem is to calculate the probability that the particle will still be on any one of the diabatic curves after a time $t$. We write the probability amplitude as
\begin{equation}
\label{3}\Psi (x,t)=\left(
\begin{array}{c}
\psi _1(x,t) \\
\psi _2(x,t)
\end{array}
\right) ,
\end{equation}
where $\psi _1(x,t)$ and $\psi _2(x,t)$ are the probability amplitude for the two states. $\Psi (x,t)$ obey the time dependent Schr$\stackrel{..}{o}$ dinger equation (we take $\hbar =1$ here and in subsequent calculations)
\begin{equation}
\label{4}i\frac{\partial \Psi (x,t)}{\partial t}=H\Psi (x,t).
\end{equation}
$H$ is defined by
\begin{equation}
\label{5}H=\left(
\begin{array}{cc}
H_1(x) & V(x) \\
V(x) & H_2(x)
\end{array}
\right) ,
\end{equation}
where $H_i(x)$ is
\begin{equation}
\label{6}H_i(x)=-\frac 1{2m}\frac{\partial ^2}{\partial x^2}+V_i(x).
\end{equation}
We find it convenient to define the half Fourier Transform $\overline{\Psi }%
(\omega )$ of $\Psi (t)$ by
\begin{equation}
\label{9}\overline{\Psi }(\omega )=\int_0^\infty \Psi (t)e^{i\omega t}dt.
\end{equation}
Half Fourier transformation of Eq. (\ref{4}) leads to
\begin{equation}
\label{11}\overline{\Psi }(\omega )=iG(\omega )\Psi (0),
\end{equation}
where $G(\omega )$ is defined by 
\begin{equation}
(\omega -H) G(\omega )=I. 
\end{equation}
In the position representation, the above equation may be written as
\begin{equation}
\label{12}\overline{\Psi }(x,\omega )=i\int_{-\infty }^\infty G(x,x_0;\omega
)\overline{\Psi }(x_0,\omega )dx_0,
\end{equation}
where $G(x,x_0;\omega )$ is
\begin{equation}
\label{13}G(x,x_0;\omega )=\langle x|(\omega -H)^{-1}|x_0\rangle .
\end{equation}
Writing
\begin{equation}
\label{14}G(x,x_0;\omega )=\left(
\begin{array}{cc}
G_{11}^{}(x,x_0;\omega ) & G_{12}^{}(x,x_0;\omega ) \\
G_{21}^{}(x,x_0;\omega ) & G_{22}^{}(x,x_0;\omega )
\end{array}
\right)
\begin{array}{cc}
&  \\
&
\end{array}
\end{equation}
and using the partitioning technique \cite{Lowdin} we can write
\begin{equation}
\label{15}G_{11}^{}(x,x_0;\omega )=\langle x|[\omega -H_1-V(\omega
-H_2)^{-1}V]^{-1}|x_0\rangle.
\end{equation}
The above equation is true for any general $V$. This expression
simplify considerably if $V$ is a delta function located at $x_c$.
 \begin{equation}
\label{22} G_{11}(x,x_0;\omega )=
G_1^0(x,x_0;\omega)+\frac{K_0^2G_1^0(x,x_c;\omega)G_2^0(x_c,x_c;\omega
)G_1^0(x_c,x_0;\omega )}{ 1-K_0^2G_1^0(x_c,x_c;\omega
)G_2^0(x_c,x_c;\omega )},
\end{equation}
where 
\begin{equation}
\label{17}G_{i}^0(x,x_0;\omega )=\langle x|(\omega -H_{i}^{})^{-1}|x_0\rangle ,
\end{equation}
and corresponds to propagation of the particle starting at $x_0$ on the second diabatic curve, in the absence of coupling to the first diabatic curve. Using the same procedure one can get
\begin{equation}
\label{23}
\begin{array}{c}
G_{12}^{}(x,x_0;\omega )=\frac{K_0G_1^0(x,x_c;\omega )G_2^0(x_c,x_0;\omega )%
}{1-K_0^2G_1^0(x_c,x_c;\omega )G_2^0(x_c,x_c;\omega )}.
\end{array}
\end{equation}
Similarly one can derive expressions for $G_{22}^{}(x,x_0;\omega )$ and $%
G_{21}^{}(x,x_0;\omega )$. Using these expressions for the Green's
function in Eq. (\ref{11}) we can calculate $\overline{\Psi
}(\omega )$ explicitly.
\newline
The expressions that we have obtained for $\overline{\Psi }(\omega)$ are quite general and are valid for any $V_1(x)$ and $V_2(x)$. However, their utility is limited by the fact that one must know $G_1^0(x,x_0;\omega )$ and $G_2^0(x,x_0;\omega )$. It is possible to find $G_i^0(x,x_0;\omega )$ only in a few limited cases and the Morse oscillator is one of them \cite{Grosche,Lawande,Duru}.
\begin{figure} \centering
\epsfig{file=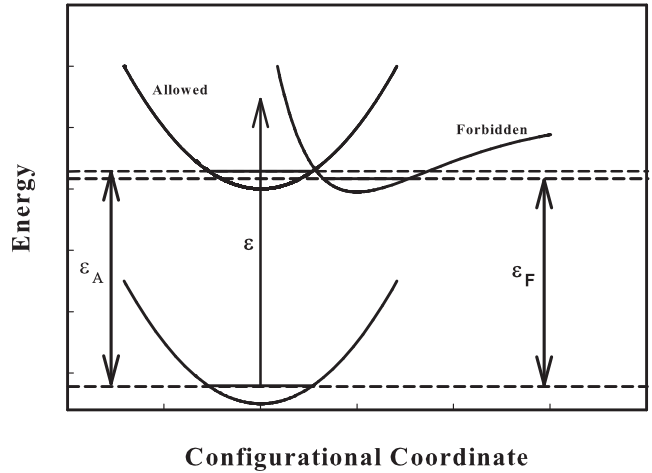,width=0.5\linewidth} 
\caption{Schematic diabatic potential energy curves
illustrating the model.} \label{Curveapply}
\end{figure}

\section{Electronic Absorption Spectra and Resonance Raman Excitation Profile : Curve Crossing induced Dissociation} 

\noindent In this section we apply the method to the problem involving a Harmonic oscillator  and a Morse oscillator, coupled by Dirac delta function. We consider a system of three potential energy curves, ground electronic state and two `crossing' excited electronic states (electronic transition to one of them is assumed to be dipole forbidden and while it is allowed to the other) \cite{Zink,ZinkPRL}. We calculate the effect of `crossing' induced dissociation on electronic absorption spectra and on resonance Raman excitation profile. The propagating wave functions on the excited state potential energy curves are given by solution of the time dependent Schr\"{o}dinger equation
%\begin{widetext}
\begin{equation}
\label{N33}i\frac \partial {\partial t}\left(
\begin{array}{c}
\psi _1^{vib}(x,t) \\
\psi _2^{vib}(x,t)
\end{array}
\right)=\left(
\begin{array}{cc}
H_{vib,e1}(x) & V_{12}(x) \\
V_{21}(x) & H_{vib,e2}(x)
\end{array}
\right) \left(
\begin{array}{c}
\psi _1^{vib}(x,t) \\
\psi _2^{vib}(x,t)
\end{array}
\right).
\end{equation}
%\end{widetext}
In the above equation $H_{vib,e1}(x)$ and $H_{vib,e2}(x)$ describes the vibrational motion of the system in the first electronic excited state (allowed) and second electronic excited state (forbidden) respectively
\begin{equation}
\label{N34}H_{vib,e1}(x)=-\frac 1{2m}\frac{\partial ^2}{\partial
x^2}+\frac{1}{2}m \omega_{A}^2(x-a)^2
\end{equation}
and
\begin{equation}
\label{N35}H_{vib,e2}(x)=-\frac 1{2m}\frac{\partial ^2}{\partial
x^2}+D_{F}[1-e^{(x-b)}]^2.
\end{equation}
In the above $m$ is the oscillator's mass, $\omega_{A}$ is the vibrational frequency of the first electronic excited state and $D_{F}$ is the dissociation energies of the  forbidden states and $x$ is the vibrational coordinate. Shifts of the vibrational coordinate minimum upon excitation are given by  $a$ and $b$, and $V_{12}$ ($V_{21}$) represent coupling between the two diabatic potentials which is taken to be
\begin{equation}
\label{N36}V_{21}(x)=V_{12}(x)=K_0\delta (x-x_c),
\end{equation}
where $K_0$ represent the strength of the coupling.
\begin{figure} \centering
\epsfig{file=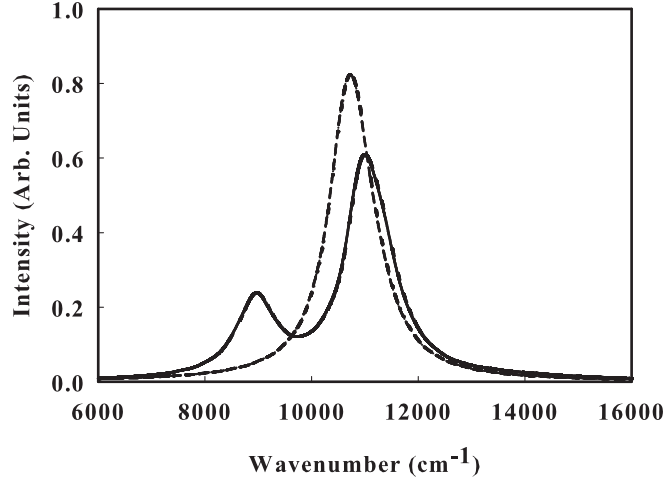,width=0.5\linewidth} 
\caption{Calculated electronic absorption spectra with coupling (solid line) and without coupling (dashed line).} \label{elec}
\end{figure}
The intensity of electronic absorption spectra is given by
\cite{Zink,Heller}
\begin{eqnarray}
\label{N41}I_A(\omega )\propto & Re[\int_{-\infty }^\infty
dx\int_{-\infty }^\infty dx_0\Psi_i ^{vib*^{}}(x)\nonumber
\\ & iG(x,x_0;\omega +i\Gamma )\Psi_i ^{vib}(x_0)],
\end{eqnarray}
where
\begin{equation}
\label{N42}G(x,x_0;\omega +i\Gamma )=\langle x|[(\omega_0/2+\omega
-\omega _{eg})+i\Gamma -H_{vib,e}]^{-1}|x_0\rangle .
\end{equation}
and
\begin{equation}
\label{N42a} H_{vib,e}=\left(
\begin{array}{cc}
H_{vib,e1}(x) & K_0 |x_c\rangle\langle x_c| \\
K_0  |x_c\rangle\langle x_c| & H_{vib,e2}(x)
\end{array}
\right)
\end{equation}
Here, $\Gamma$ is a phenomenological damping constant which account for the life time effects. $\Psi_i^{vib}(x,0)$ is given by
\begin{equation}
\label{N42b}\Psi _i^{vib}(x,0)=\left(
\begin{array}{c}
\chi_i(x) \\
0
\end{array}
\right),
\end{equation}
where $\chi_i(x)$ is the ground vibrational state of the ground electronic state, $\omega_0$ is the vibrational frequency on the ground electronic state, $\varepsilon_A$ is the energy difference between the excited (allowed) and ground electronic state, and for the forbidden electronic state it's value is $\varepsilon_F$. Similarly resonance Raman scattering intensity can be expressed in terms of Green's function and is given by \cite{Heller,Zink}.
\begin{eqnarray}
\label{N53} I_R(\omega )\propto &|\int_{-\infty }^\infty
dx\int_{-\infty
}^\infty dx_0\Psi _f^{vib*}(x,0)\nonumber\\
& iG(x,x_0;\omega+i\Gamma)\Psi _i^{vib}(x_0,0)|^2.
\end{eqnarray}
In the above $\Psi _f^{vib}(x,0)$ is given by
\begin{equation}
\label{N53a}\Psi _f^{vib}(x,0)=\left(
\begin{array}{c}
\chi _f(x) \\
0
\end{array}
\right),
\end{equation}
where $\chi _f(x)$ is the final vibrational state of the ground electronic state. As $G_i^0(x,x_0;\omega)$ for the harmonic potential is known \cite{Grosche}, we can calculate
$G(x,x_0;\omega)$. We use Eq. (\ref{N53}) to calculate the effect of curve crossing induced dissociation on resonance Raman excitation profile.

\subsection{Results using the model}

\begin{figure} \centering
\epsfig{file=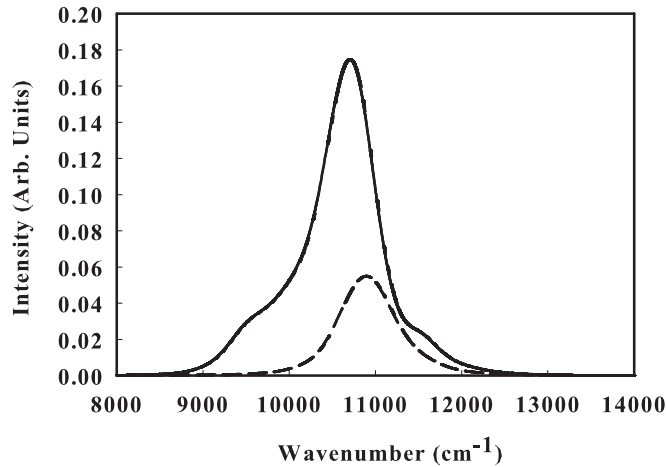,width=0.5\linewidth} 
\caption{Calculated resonance Raman excitation profile for excitation from the ground vibrational state to the first excited
vibrational state, with coupling (solid line) and without coupling (dashed line).} \label{raman}
\end{figure}

\noindent In the following we give results for the effect of curve crossing on electronic absorption spectrum and resonance Raman excitation profile in the case where one dipole allowed electronic state
crosses with a dipole forbidden electronic state as in Fig. \ref{Curveapply}. As in \cite{Zink}, the ground state curve is taken to be a harmonic potential energy curve with its minimum at
zero. The curve is constructed to be representative of the potential energy along a metal-ligand stretching coordinate. We take the mass as $35.4\:amu$ and the vibrational wavenumber as
$400\:cm^{-1}$ \cite{Zink} for the ground state. The first diabatic excited state potential energy curve is displaced by $0.1\:\AA$ and is taken to have a vibrational wavenumber of $400\:cm^{-1}$. Transition to this state is allowed. The minimum of the potential energy curve is taken to be above $10700\:cm^{-1}$ of that of the ground state curve. The second diabatic excited state potential energy curve is taken to be an un-displaced excited state. On that potential energy curve, the vibration is taken to have same wavenumber of $400\:cm^{-1}$. Its minimum is $10800\:cm^{-1}$ above that of the ground state curve.
Transition to this state is assumed to be dipole forbidden. The two diabatic curves cross at an energy of $10804.1\:cm^{-1}$ with $x_c=-0.02477\:\AA$. Value of $K_0$ we use in our calculation is $K_0=5.54275 \times 10^{-15}\:erg.\AA$. The lifetime of both the excited states are taken to be $450\:cm^{-1}$. The calculated electronic absorption spectra is shown in Fig. \ref{elec}. The profile shown by the dashed line is in the absence of any coupling to the second potential energy curve. The full line has the effect of coupling in it. The calculated resonance Raman excitation profile is shown in Fig. \ref{raman}. The profile shown by the full line is calculated for the coupled potential energy curves. The profile shown by the dashed line is calculated for the uncoupled potential energy curves. It is seen that curve crossing effect can alter the absorption and Raman excitation profile significantly. However it is the Raman excitation profile that is more effected.

\section{Conclusions}

\noindent In our earlier paper we have proposed an exactly solvable model for the two state curve crossing problem. In this paper we have extended our model to deal with the case of curve crossing induced dissociation. We have analyzed the effect of curve crossing on electronic absorption spectrum and Resonance Raman excitation profile for a model consisting of a Harmonic oscillator  and a Morse oscillator, coupled by Dirac delta function. We find that the Raman excitation profile is affected much more by the crossing than the electronic absorption spectrum. The same procedure is also applicable to the case where $S$ is a non-local operator, and may be represented by $S = |f> K_0 <g|$, $f$ and $g$ are arbitrary acceptable functions. Choosing both of them to be Gaussian will be an improvement over the delta function coupling model, $S$ can also be a linear combination of such operators. 

\section{Acknowledgments}

\noindent The author thanks Prof. K. L. Sebastian for valuable suggestions. It is a pleasure to thank Prof. M. S. Child for his kind interest, suggestions and encouragement. The author thanks Prof E. E. Nikitin and Prof. H. Nakamura for sending helpful reprint of their papers. The author also thanks Prof. H. Klienert for his useful comment on Green's function of Morse oscillator.

\end{document}